\documentclass[journal]{IEEEtran}
\ifCLASSINFOpdf
\else
\fi
\usepackage{subfigure}
\usepackage{multirow}
\usepackage{url}
\usepackage{tabularx}
\usepackage{booktabs}
\usepackage{multicol}
\usepackage{afterpage}
\usepackage{amsmath}
\usepackage{float}
\usepackage{graphicx}
\usepackage{xcolor}
\usepackage{authblk}

\begin{document}
%
\title{SCALHEALTH: Scalable Blockchain Integration for Secure IoT Healthcare Systems}
%
%
%

\author{Mehrzad Mohammadi,
        Reza Javan, Mohammad Beheshti-Atashgah
        and~Mohammad Reza Aref 
\thanks{The authors are with the Information Systems and Security Laboratory,
Department of Electrical Engineering, Sharif University of Technology,
Tehran 14588-89694, Iran. E-mail: mohammadimehrzad11@gmail.com, reza.javan@alum.sharif.edu, Mohammad.beheshti@researcher.sharif.edu, aref@sharif.edu.}
\thanks{}
\thanks{}}

%
%

\markboth{}%
{Shell \MakeLowercase{\textit{et al.}}: SCALHEALTH: Scalable Blockchain Integration for Secure IoT Healthcare Systems}
%



\maketitle

\begin{abstract}
Internet of Things (IoT) devices are capable of allowing for far-reaching access to and evaluation of patient data to monitor health and diagnose from a distance. An electronic healthcare system that checks patient data, prepares medicines and provides financial assistance is necessary. Providing safe data transmission, monitoring, decentralization, preserving patient privacy, and maintaining confidentiality are essential to an electronic healthcare system. In this study, we introduce (SCALHEALTH) which is a blockchain-based scheme of the Hyperledger Fabric consortium. In this study, we use authentication to agree on a common key for data encryption to send data confidentially. Also, sending data through IPFS is decentralized. Non-fungible token (NFT) is used to send patient prescriptions to pharmacies and insurance companies to ensure the authenticity of patient prescriptions. As the system's main body, blockchain creates authorization and validation for all devices and institutions. Also, all metadata in the system is recorded on the blockchain to maintain integrity, transparency, and timely data monitoring. The proposed study uses two types of blockchain: a health blockchain and a financial blockchain. The financial blockchain is for financial transactions and is based on Ethereum. The health blockchain also introduces a mechanism that allows several blockchains to be active in parallel, instead of only one blockchain. The prototype of this mechanism is simulated in two scenarios. In comparison to the normal state, the proposed plan has superior results.
\end{abstract}


%
\IEEEpeerreviewmaketitle

\section{Introduction}
%
%
%
%
\IEEEPARstart{T}he Internet of Things (IoT) has transformed healthcare by allowing extensive volumes of medical information to be stored and transmitted by IoT networks [1]. Yet, security and privacy are still major concerns. Healthcare systems must ensure that their protection protocols encompass secure transmission, tamper-proof monitoring, data storage, integrity, and privacy-aware authentication [2], [3]. In the healthcare sector, access control, provenance, data integrity, and interoperability are critical to maintaining patient privacy and enabling data exchange with other institutions [4]. Data integrity deals with data quality, which is critical for healthcare institutions currently experiencing an increasing demand for real-world data from industry and research organizations [5]. However, unauthorized sharing, public theft, and theft of sensitive information erode public trust in healthcare institutions. Thus, blockchain technology can potentially improve access control, interoperability, provenance, and data integrity while establishing and maintaining trust among stakeholders [6]. In this regard, Blockchain technology presents a promising option for improving IoT healthcare systems' security and privacy levels [7]. Blockchain's immutable feature enables e-health data to be stored securely at medical blocks to assert data privacy and integrity [8], [2], [9]. Additionally, blockchain's decentralized nature could potentially provide a level of security without relying on intermediaries [10].

In this article, we propose an electronic health system framework based on a consortium blockchain in which patient data is sent securely and seamlessly. We also introduce a monitoring mechanism for all devices in the electronic health system. This will prevent the activity of unauthorized devices and prevent false information from being injected into the system.  The patient and the hospital agree on a shared key through an authentication protocol. Encryption and confidentiality of patient data are done through this key. We establish a reliable relationship between the hospital, pharmacy, and insurance. This is so that the diagnosis of real patients can be done by insurance and pharmacy for the preparation of medicine and the costs of treatment assistance. For better scalability of the system, we offer a mechanism for the concurrent operation of several blockchains. Using simulation results, we show the superiority of the concurrent activity of several blockchains.

\section{BACKGROUND}
\subsection{Blockchain}
Blockchains are tamper evident and tamper resistant digital ledgers implemented in a distributed fashion (i.e., without a central repository) and usually without a central authority (i.e., a bank, company or government). At their basic level, they enable a community of users to record transactions in a shared ledger within that community, such that under normal operation of the blockchain network no transaction can be changed once published [11].
Blockchain is divided into three major categories based on its application, including 1- Public blockchain 2- Private blockchain 3- Consortium or semi-private blockchain.
\subsubsection{Public Blockchain}  In an unauthorized or public blockchain, participants in the system do not need permission to join the network [12]. This blockchain is truly decentralized because participants can participate in the consensus process, read and send transactions, and maintain a shared ledger [13]. New formats can be published, accessed, and validated by all participants, so they can keep a version of the complete blockchain [14].
\subsubsection{Private Blockchain}
Authorized or private blockchains are designed for an organization. Participants are allowed to join the network by invitation and play a specific role in maintaining the blockchain in a decentralized manner [12].
\subsubsection{Consortium Blockchain} Consortium blockchains are similar to private blockchains, but they are designed for multiple organizations. Only invited and trusted participants are allowed to join and maintain the network [15].

A blockchain consists of the following main components: 1- Transactions: A transaction is a unique piece of information, action, or exchange intended to be stored on the blockchain. 2- Blocks: A block consists of a collection of valid transactions. A block also contains necessary metadata, such as a timestamp, hash, and information about the recorded data. 3- Miners: Participants in the network who are responsible for creating blocks. A reward is given to the miner for each valid block that is created.4- Consensus: An agreement among network participants about the state of a blockchain is called consensus. Multiple nodes may be involved in a blockchain network that can create and validate transactions. The nodes, however, must agree on the order and content of the transactions that will be added to the blockchain.5- Smart Contract: A smart contract is a self-executing program stored on a blockchain that automatically executes transactions based on predetermined conditions, eliminating the need for intermediaries and enabling immediate confirmation between all parties.


\subsection{Ethereum }
In 2014, Vitalik Buterin published an article [16] referring to deficiencies in the Bitcoin blockchain, including low transaction throughput or poor scalability, significant delays in creating a block in Bitcoin, and the absence of necessary practical tools such as smart contracts for Bitcoin development. Bitcoin deficiencies led Vitalik Buterin to develop Ethereum. This allowed developers to build consensus-based applications with scalability, standardization, completeness, ease of development, and expansion capabilities. Ethereum allows anyone to write smart contracts and decentralized applications in which individuals can create their own rules for owning various transactions.
\subsection{Hyperledger Fabric}
Hyperledger Fabric is an open-source, modular, consortium blockchain platform designed by IBM for developing applications and solutions in organizational contexts. It offers plug-and-play components, such as consensus and membership services, and utilizes containers to host smart contracts or chaincodes, which contain the system's application logic. Hyperledger Fabric also offers unique features that distinguish it from other distributed ledger or blockchain applications [17-18].
\subsection{NFT}
Non-fungible token (NFT): The purpose of these signs is to indicate unique ownership. These signs mark things like real estate to confirm ownership. These tokens have only one owner and cannot be managed by other people or copied, and the security of these tokens is guaranteed by the blockchain structure. First, it should be known that these signs are not completely unique and divisible. Any piece of data can be marked with these tokens to identify and claim ownership of that data, and all of these tokens can be traced. NFTs can store many things such as video, music, legal documents, invoices, and many more. A valid identifier and metadata are used to manage mark ownership, which cannot be changed by other NFTs. NFTs are concluded through smart contracts that assign ownership and manage the transferability of the NFT. With NFTs, we are provided with the following features: the validity of NFTs is recorded on the Blockchain and prevents fraud and manipulation, and ownership of NFTs is transparent. The uniqueness of an NFT on a blockchain allows for tracing its owner, proving its legitimacy of the NFTs and irrefutable ownership. Another positive aspect of these symptoms is their transferability [19, 20].
\subsection{IPFS}
InterPlanetary File System (IPFS) is a protocol and network designed to create a permanent and decentralized method of storing and sharing files on the internet. Rather than relying on a centralized server to host files, IPFS uses a peer-to-peer network where each participating node stores a copy of the file, allowing for faster and more reliable access to content. This decentralized approach also provides greater security and privacy, as files are not controlled by a single entity and cannot be censored or taken down by a single point of failure. IPFS achieves its decentralized and distributed nature through a content-addressed system. Files are stored and retrieved using their unique cryptographic hash, which serves as their identifier. This hash can be used to access the file from any node in the network, without needing to know its location or be concerned with the underlying infrastructure. IPFS also includes built-in versioning, enabling users to access previous versions of files and ensuring that files cannot be overwritten or tampered with without detection [21-22].
\section{RELATED WORK}
Bitcoin has features such as transparency, security, immutability of recorded data, and decentralization using blockchain technology. These features have led researchers to pay special attention to this area for other systems. Research by A. Dorri and his colleagues, who use blockchain in the Internet of Things, is mentioned [23-24]. In this research, a decentralized network based on cluster networks was introduced. This network decentralizes and ensures data integrity, confidentiality, and privacy. One of the problems with this research was the lack of attention to motivational issues. Based on these studies, other research was conducted by A. D. Dwivedi and his colleagues [25]. This research presented a more advanced security and privacy model for the electronic health system that included necessary sensitivities in the health system. Due to these sensitivities, a separate system called the Internet of Medical Things (IOMT) was developed. Some networks for registering and transferring medical data that use blockchain technology use non-centralized financial networks, which can inherit the negative features of non-centralized financial networks in their structure. For example, Bitcoin, due to the use of the UTXO structure for financial transactions and the ability to establish connections between transactions, does not have complete anonymity [26], so it is natural that networks that use Bitcoin also do not have complete anonymity.
Ethereum has less delay than Bitcoin and also provides smart contracts. Therefore, using the Ethereum network is more appropriate. The first idea for registering and transferring electronic medical records (EMR) using smart contracts on Ethereum for registering medical data and limiting access to this data was published in 2015 by A. Azaria and colleagues [27], but this idea had problems such as not paying attention to patients' privacy and having a central server that would cause disruption to the system if there were any problems with the central server. Given that some research did not address when we could use blockchains, in 2018 an article was presented under the same title [28]. This article provided a complete understanding of blockchains. In the same year, a direct search plan through encrypted data was presented by Shengshan Hu and colleagues [29]. In this article, a searchable system based on smart contracts was presented. This system restricted people's access to data and prevented privacy breaches. To complete it, in 2019, to prevent data leakage in electronic health records (EHR) that violates privacy, a searchable encryption plan based on blockchains was proposed by Lanxiang Chen and colleagues [30]. In this article, a searchable encryption plan based on blockchains was proposed for EHR. In 2020, a blockchain-based healthcare system with smart contract capabilities was presented by MANAF ZGHAIBEH and colleagues [31].
In this article, a system called SHealth is introduced. This is a private chain of templates hosted and maintained by various institutions, including the health management system. This private chain of templates is connected to a distributed ledger database and contains records of all users. It is safely accessible to all system entities and nodes. In 2021, a plan for electronic health record services in the IoT-Blockchain system was introduced by Partha Pratim Ray and colleagues. This article proposes a new privacy preservation plan based on blockchain methods and congestion exchange. This plan is to facilitate user data transfer without any problems or risks [32]. Some articles also focus on how to create an electronic health network that is highly efficient in terms of energy, information transmission volume, and memory requirements. This is so that the newly created network can be operational [33-34].
\section{PROPOSED SCHEME}
In this section, we first explain all the stages of the Electronic Health System mechanism. Then we describe the system's blockchain structure.
Assuming that all devices, individuals, and institutions within the system have been registered for activities, we now examine the steps of the system's mechanism. \subsection{Steps for Analyzing patient data}
\subsubsection{Phase zero}
First, we introduce the Organizational Components of the Proposed Health System.
Patients: Individuals enrolled in the proposed health system to maintain their health.
Hospitals: Trusted institutions where patients receive healthcare services. Hospitals consist of doctors, nurses, laboratories, and other healthcare professionals who monitor and maintain their patients' health. It is worth mentioning that every hospital has local storage for storing the essential data of the hospital and its patients.
Insurance Companies: Partners of the patients responsible for covering some treatment costs.
Pharmacies: Supply medicines to patients within the proposed health system.
Device Manufacturers: Organizations responsible for overseeing patient wearable sensors and patient/hospital gateways within the system. They also update these devices and prevent unauthorized devices from entering the system.
\subsubsection{Phase one}
All data related to the patient's wearable sensors are collected on the patient's mobile device.
\subsubsection{Phase two}
Implementation of the authentication protocol is as follows.
An authentication protocol between the patient and the hospital is implemented in this article, whose security requirements have been established[35]. Table 1.3 illustrates the symbols used in this protocol before describing the steps.
First, we assume that each hospital gateway is a trusted party and that every message is transmitted through the hospital gateways marked with $\text{GID}_j$ in the protocol. To initialize the network, each gateway generates a secret key called $\text{G}_j$ and selects a collision-resistant cryptographic hashing function $H$. Second, the registration stage: the registration stage includes steps for registering users on the portal. In this study, all the sensors attached to the patient are assumed to be recorded on the patient's phone.\par This stage involves three steps. \begin{enumerate} 
\item \textbf{Step 1:} $P_i$ with identity $MID_i$ chooses a password $PW_i$ to generate $H(MID_i \Vert PW_i)$ and also chooses a secret random $N_0$. Then it sends $MID_i$ and $H(MID_i \Vert PW_i)$ through a secure channel to the gateway of hospital $j$.
\item \textbf{Step 2:} If $MID_i$ is not already registered, the $j$th hospital gateway creates a temporary alias $CID_i$ and a random number $R_x$ for $P_i$ and stores them with $MID_i$ and $H(MID_i \Vert PW_i)$ in its database, then calculates $A_1$ and $A_2$ and sends $GID_j$, $CID_i$, $A_1$, and $A_2$ to $U_i$ through a secure channel.
\item \textbf{Step 3:} In this step, $P_i$ calculates $A_3$ relations of $P_i$ and saves $A_1$, $A_2$, $A_3$, $CID_i$, and $GID_j$ in his phone. Now $P_i$ is ready to authenticate on an insecure channel.\end{enumerate} This is shown in Figure 3.3.
\begin{equation}
\label{eq:formula1}
A_1 = H(CID_i \Vert Rx \Vert GID_j \Vert Gj) \oplus H(MID_i \Vert PW_i)
\end{equation}

\begin{equation}
\label{eq:formula2}
A_2 = H(MID_i \Vert GID_j) \oplus H(MID_i \Vert H(MID_i \Vert PW_i)
\end{equation}

\begin{equation}
\label{eq:formula3}
A_3 = H(MID_i \Vert PW_i) \oplus N_0
\end{equation}

\begin{table}[htbp]
  \centering
  \caption{Introducing the symbols of the authentication algorithm}
  \label{tab:authsymbols}
    \begin{tabular}{|c|c|}
    \hline
    Symbol & Description \\
    \hline
    $P_i$ & $i^{\text{th}}$ patient ID \\
    $MID_i$ & Mobile ID of $i^{\text{th}}$ patient  \\
    $H(\cdot)$ & Cryptographic hash function \\
    $\Vert$ & Concatenation operator \\
    $\oplus$ & Bitwise XOR operator \\
    $PW_i$ & The password of $i^{\text{th}}$ patient  \\
    $Sk_p$ & Common session key generated by the patient \\
    $Sk_g$ & Common session key generated by the hospital \\
    $M_i$ & messages in authentication step \\
    $CID_i$ & Patient’s temporary pseudonyms \\
    $R_p,R_g,R_x$ & temporary random numbers \\
    $GID_i$ & Hospital Gateway ID of $i^{\text{th}}$ \\
    $N_0$ & Nonce \\
    \hline
    \end{tabular}%
\end{table}%

\begin{figure}[htbp]
    \centering
    \includegraphics[width=3in]{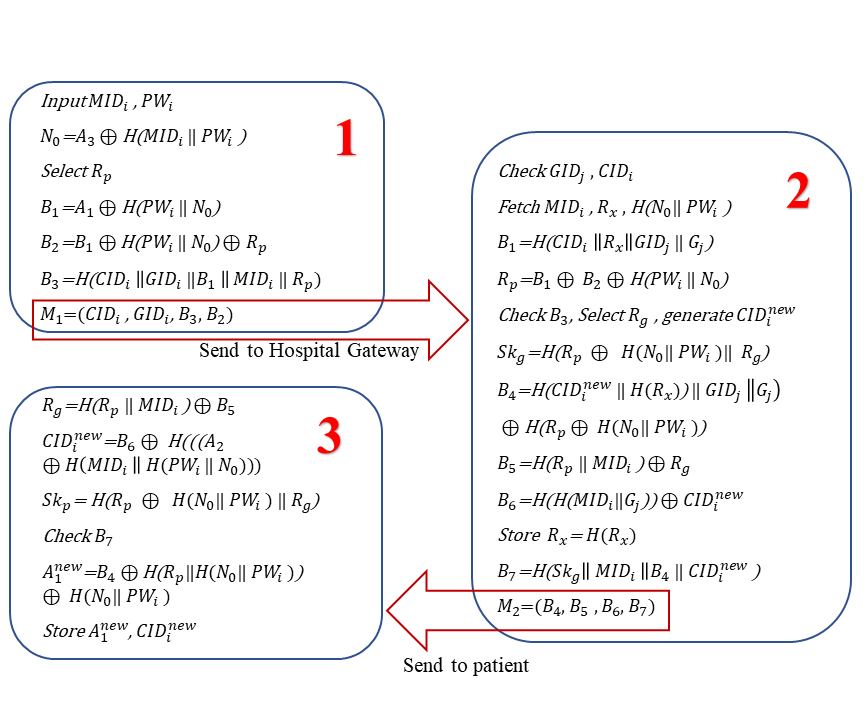}
    \caption{Proposed Authentication algorithm}
    \label{fig:example}
\end{figure}

After the patient's authenticity has been verified, the patient can communicate with the hospital over an insecure channel. This will enable the patient to establish a symmetric shared key with the hospital for the purpose of encrypting data.
the patient encrypts his data using the symmetric encryption key agreed upon in the authentication protocol and transmits the encrypted data to the hospital. Using symmetric encryption can have significant advantages, such as speeding up the encryption and decryption of patient and hospital data [36].
\subsubsection{Phase three}
The hospital decrypts patients' data. We propose a system for the secure transmission of patients' medical data and prescriptions from hospitals to pharmacies and insurance companies. This is done while ensuring authenticity and preventing manipulation.

We assume that each hospital has a supervised machine learning model, which categorizes patients' data into critical and non-critical parts. Doctors examine critical data first, and numerous prescriptions are created daily.

To ensure the integrity of the prescriptions, we propose a mechanism that verifies the authenticity of the message and the source of the prescription. This mechanism will enable pharmacies to receive prescriptions from doctors without any manipulation. Additionally, it will also provide insurance companies with access to prescriptions written by their policyholders. By having access to this information, the insurance company can cover its patients without any ambiguity.
In this study, we utilize NFTs to collect prescriptions created by doctors in hospitals over a certain period of time. Prescription metadata is stored in the NFT, and all prescription data is encrypted and added to the IPFS. A link is then created and placed in the relevant NFT, which is sent to the pharmacy and insurance companies.

Upon receiving the relevant NFT, the pharmacy and insurance company can verify the validity and integrity of the information through the unique characteristics of the NFT. The attached prescription data is then used by the pharmacy to dispense medication to patients, and the related blockchain transactions are recorded promptly.

This system not only enables accurate tracking of patient treatment costs but also provides necessary support for each patient through their insurance. Additionally, the pharmacy can prioritize the preparation of medications for patients with critical conditions, ensuring faster access to treatment.
Our proposed system ensures a safe and efficient way to transfer patients' medical data and prescriptions. This will benefit hospitals, doctors, pharmacies, and insurance companies.
\subsection{Proposed blockchain structure}

The proposed health system includes several technology components, which aim to enhance the system's efficiency and security. One of the key components is the system blockchain, which consists of two main blockchains: financial and health. The financial blockchain is built on the Ethereum platform and is used for all financial transactions within the system. This includes exchanges between patients, hospitals, pharmacies, and insurance companies.
The health blockchain, on the other hand, is a consortium blockchain based on Hyperledger Fabric. It is designed to record and store all health-related metadata in the system. The health blockchain miners are divided into three categories: hospital miners, insurance miners, and device manufacturer miners.
Hospital miners record all health-related metadata, including patient information, hospital records, pharmacy records, and insurance records. Anyone with the necessary qualifications can become a hospital miner, but they must be accredited by the health system's accredited body.

Insurance miners, on the other hand, are installed by insurance companies. They are responsible for registering people covered by insurance, updating the list of people covered, and verifying the registered metadata related to insurance.

Device manufacturer miners are identified by device manufacturers and verify the unique identifiers of devices in the system. They also update the programs of authorized devices in the system, issue licenses for newly installed devices, and block unauthorized devices.

Overall, the proposed technology components of the health system, including the blockchain infrastructure, aim to improve the security, transparency, and efficiency of the system. This will result in better healthcare outcomes for patients.

In financial systems, transactions are interdependent. There is a need for a single blockchain to record all financial transactions in order to ensure that they cannot be tampered with. But, healthcare systems differ from financial systems in that patient transactions are independent of each other. Each patient's data is specific to them and has no relationship with other patients. In the proposed healthcare system, each hospital will have its own blockchain to record patients' transactions specific to that hospital. Relevant metadata will be recorded on each hospital's blockchain, which can be shared with other hospitals on the network.
In Section V, we demonstrate the advantages of each hospital recording patient transactions on its own blockchain.
\subsection{The Connection Between the Blockchain of Hospitals with Each Other and the Financial Blockchain}
Data exchange between hospitals has become increasingly important in healthcare. In cases where a patient needs to move from one hospital to another, the current hospital needs to receive information from the previous hospital to ensure proper care. In this context, communication between hospitals is essential.

\begin{figure}[!b]
  \centering
  \includegraphics[width=1.0\textwidth, height=0.32\textheight]{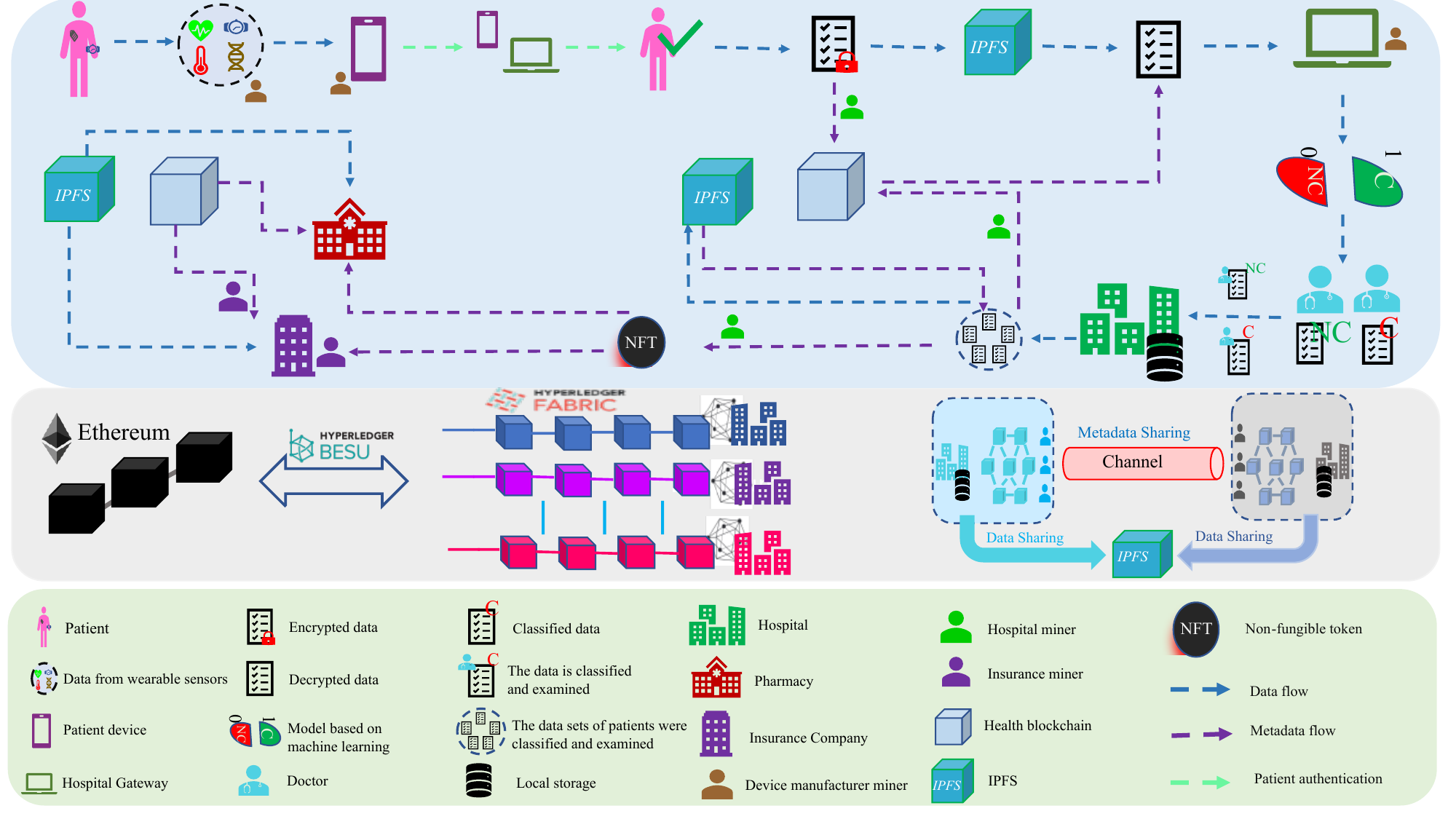}
  \caption{ An Overview of the Proposed Health System}
  \label{fig:your-label}
\end{figure}

Data exchange between hospitals occurs by creating a channel in the Hyperledger Fabric. Through this communication channel, hospitals can securely share patient data and metadata. For example, when a patient wants to move from one hospital to another, the current hospital can request permission to access the patient's data from the previous hospital's local storage. By ensuring that patient data is easily accessible to all authorized parties, hospitals can provide better care and enhance patient trust in the healthcare system. 
The communication pathway between Ethereum financial and Hyperledger-Fabric is made possible through Hyperledger-Besu, which is a technology designed to establish a connection with Ethereum financial. By utilizing Hyperledger Besu, organizations can seamlessly integrate Ethereum financial with their Hyperledger Fabric network, enabling the secure and efficient exchange of data and assets.
\section{RESULTS AND EVALUATION}
\subsection{Analysis of Blockchain structure}
We conducted simulations on the ubuntu-22.04 using a device with Intel Core i7-7500U @ 2.70-2.90 GHz and 12.GB RAM and Hyperledger fabric 2.2.1 and Caliper v0.5.0 to demonstrate the efficiency improvement of our proposed blockchain structure.

In our scenario, two hospitals with equal numbers of transactions were considered. Normally, 16 miners would record the transactions of both hospitals in a single blockchain. However, in our proposed method, each hospital has its own blockchain and 8 miners are assigned to each blockchain. Our simulations showed that using separate blockchains for each hospital greatly improves scalability and efficiency. For instance, when comparing two blockchains with the same number of miners and transactions to a single blockchain with the same workload, we observed significant improvements in transaction processing time, throughput, and resource utilization. Figure 4 provides a representation of these results.
In Figure 4, Parts (a) and (b) depict the throughput and time taken to record all transactions for the create function (which represents transaction recording in the distributed ledger) based on the number of transactions. The proposed scenario exhibits a significant advantage over the normal mode, enabling a higher number of transactions to be recorded per second and reducing the overall time required to record all transactions in the blockchain. This reduction in transaction recording time not only facilitates online operation but also enhances system efficiency.
Parts (c) and (d) of Figure 4 illustrates the throughput and time taken to read total transactions for the Query function (which involves reading transactions from the distributed ledger) based on the number of transactions. In the proposed model, whenever a new miner is selected, it can read transactions from the distributed ledger more quickly.

\newpage

\begin{figure}[htb]
  \centering
  \includegraphics[width=1.0\textwidth, height=0.3\textheight]{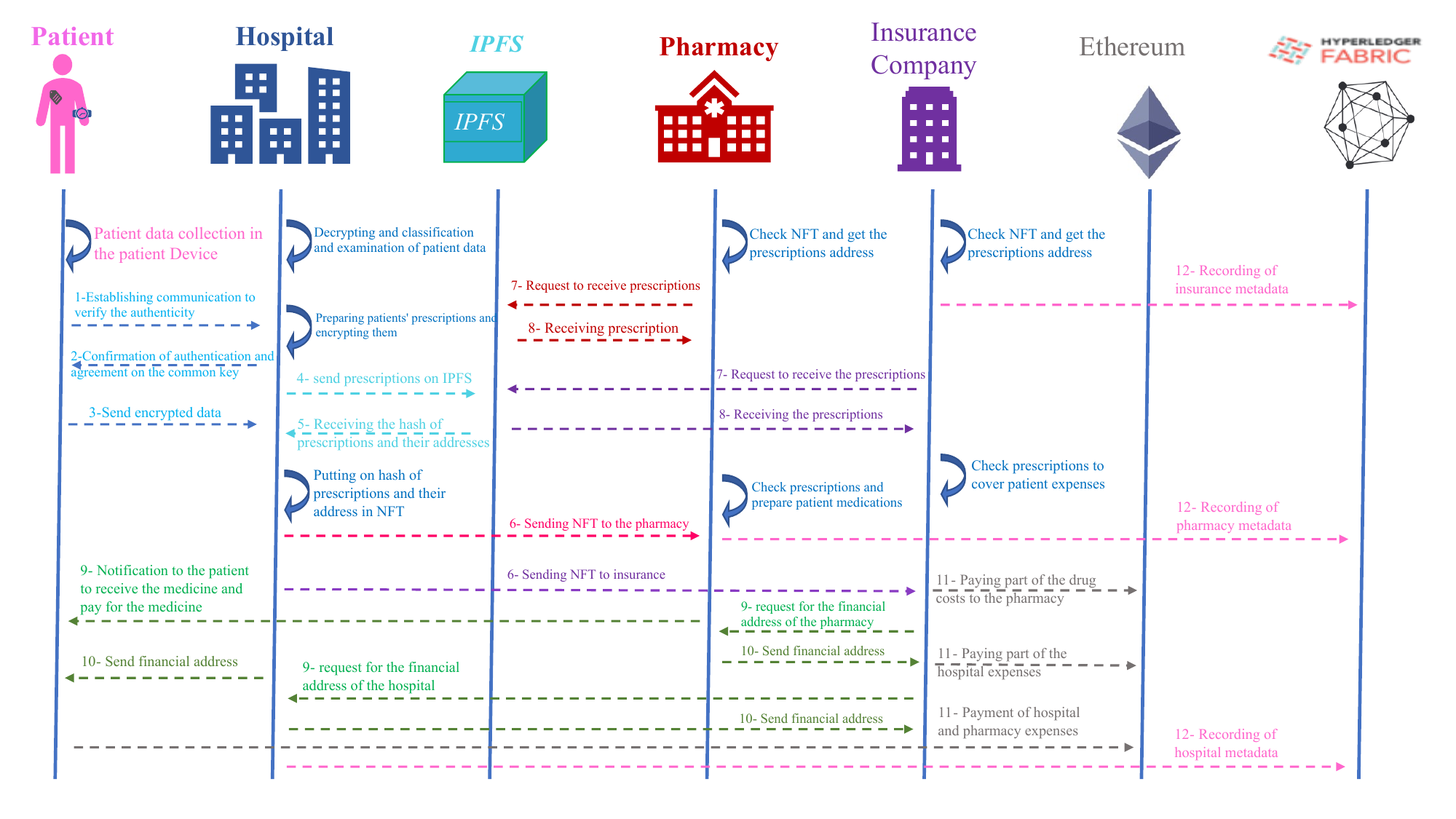}
  \caption{The overall transactions' workflow.}
  \label{fig:your-label}
\end{figure}

\begin{figure}[hbtb]
  \centering
  \includegraphics[width=1.0\textwidth, height=0.35\textheight]{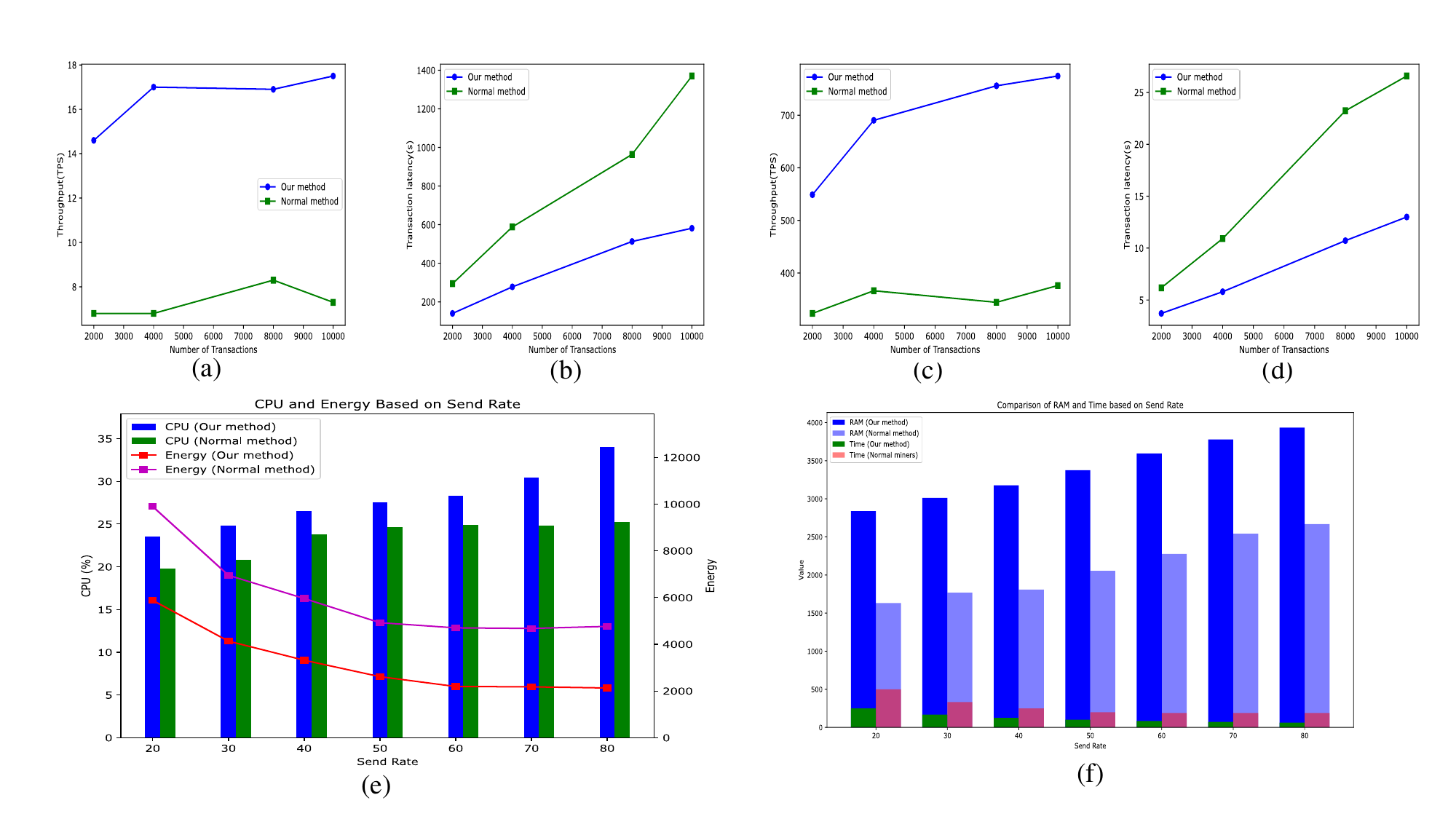}
  \caption{Simulation results for the Proposed structure.}
  \label{fig:your-label}
\end{figure}
 This enables swift access to transactions in the system for the miner.
 In part (e) of Figure 4, we present the percentage of CPU consumption and energy consumption. Comparing the CPU consumption patterns, the proposed method exhibits higher utilization compared to the normal method. However, it is important to highlight that the proposed method achieves this level of CPU consumption within a shorter time frame than the normal method. To quantify the total CPU consumption, we measured its usage and assumed that each unit of the obtained value corresponds to X units of energy consumption. Subsequently, we conducted a comparative analysis of energy consumption between the two scenarios, revealing a lower energy consumption in the proposed mode.
Additionally, Part (f) of Figure 4 provides information on the \\
\\
\vspace{17.7cm}

amount of RAM consumed in MiB and the time required to record transactions in seconds. It is evident from the figure that the proposed scenario exhibits higher RAM consumption. However, transactions are recorded in a shorter duration. For instance, at a send rate of 60 TPS in the proposed scenario, the RAM consumption is 3594.55 MiB, and the time required to record transactions is 83.34s, whereas in the normal mode, the RAM consumption is 2227.2 MiB, and the time required to record transactions is 188.68s.
\subsection{Comparison of system efficiency}
We compared our proposed system with a number of other systems, the results of which can be seen in Table 2.

\clearpage

\begin{table}[htbp]
  \centering
  \caption{Comparison of the proposed method with other works}
  \begin{tabularx}{\textwidth}{@{}l *{6}{>{\centering\arraybackslash}X}@{}}
    \toprule
    \textbf{Criteria} & \textbf{Blockchain Type} & \textbf{Blockchain Platform} & \textbf{Encryption and Decryption time} & \textbf{Off-chain Storage} & \textbf{Scalability} &  \textbf{Really decentralize} \\
    \midrule
    BIoTHR[32] & Private & - & High & Swarm Node & Low & in data:Yes, \hspace{0.7cm}in metadata:No \\
    \bottomrule
    ACTION-EHR[37] & Private & Hyperledger Fabric & High & Cloud & Medium & in data:No, \hspace{0.7cm}in metadata:No \\
    \midrule
    SHealth[31] & Private & Hyperledger Fabric &  High & - & Low & in data:-, \hspace{0.9cm}in metadata:No \\
    \midrule
    D.C. Nguyen \textit{et al.}[38] & Private & Ethereum & High & IPFS & Low & in data:Yes, \hspace{0.7cm}in metadata:No \\
    \midrule
    Hao Guo \textit{et al.}[39] &  Private & Hyperledger Fabric & High & Edge+IPFS & Low & in data:Yes, \hspace{0.7cm}in metadata:No \\
    \midrule
    Proposed Work & Consortium & Hyperledger Fabric + Ethereum & Low & Local Storage + IPFS & High & in data:Yes, \hspace{0.7cm}in metadata:Yes \\
    \bottomrule
  \end{tabularx}
\end{table}

\section{Conclusion}
This paper presents a novel framework for securely transferring and reviewing patient data. Using two types of blockchains, namely health and financial, the framework records all system transactions, providing an added level of security. In addition, we have proposed a separate health blockchain structure for storing independent transactions. Simulations and comparisons of our framework with similar systems have demonstrated its superior efficiency. The proposed solution has the potential to positively impact the healthcare industry by enabling the secure transfer and review of patient information.


%

\ifCLASSOPTIONcaptionsoff
  \newpage
\fi



%




\end{document}